\documentclass{article}

\PassOptionsToPackage{sort&compress,numbers,super}{natbib}
\usepackage[preprint]{neurips_2024}

\usepackage[utf8]{inputenc} % allow utf-8 input
\usepackage[T1]{fontenc}    % use 8-bit T1 fonts
\usepackage{url}            % simple URL typesetting
\usepackage{booktabs}       % professional-quality tables
\usepackage{amsfonts}       % blackboard math symbols
\usepackage{nicefrac}       % compact symbols for 1/2, etc.
\usepackage{microtype}   
\usepackage{graphicx}% microtypography
\usepackage{wrapfig}
\usepackage{subcaption}
\usepackage{xcolor}  
\usepackage{amsmath}% math
\usepackage[ruled, vlined]{algorithm2e}
\usepackage{algpseudocode}
\usepackage{lineno}
\let\OLDthebibliography\thebibliography
\renewcommand\thebibliography[1]{
  \OLDthebibliography{#1}
  \setlength{\parskip}{0pt}
  \setlength{\itemsep}{0pt plus 0.3ex}
}

\usepackage{hyperref}

\title{Tango*: Constrained synthesis planning using chemically informed value functions}

\author{
  Daniel Armstrong\textsuperscript{1},\quad Zlatko Jončev\textsuperscript{1},\quad Jeff Guo\textsuperscript{1,2},\quad Philippe Schwaller\textsuperscript{1,2} \\
  \textsuperscript{1}\'Ecole Polytechnique F\'{e}d\'{e}rale de Lausanne (EPFL) \\
  \textsuperscript{2}National Centre of Competence in Research (NCCR) Catalysis \\
  \texttt{\{daniel.armstrong,philippe.schwaller\}@epfl.ch} \\
}

\begin{document}

\maketitle

\begin{abstract}
  Computer-aided synthesis planning (CASP) has made significant strides in generating retrosynthetic pathways for simple molecules in a non-constrained fashion. Recent work introduces a specialised \textit{bidirectional} search algorithm with forward and retro expansion to address the starting material-constrained synthesis problem, allowing CASP systems to provide synthesis pathways from specified starting materials, such as waste products or renewable feed-stocks. In this work, we introduce a simple guided search which allows solving the starting material-constrained synthesis planning problem using an existing, uni-directional search algorithm, Retro*. We show that by optimising a single hyperparameter, Tango* outperforms existing methods in terms of efficiency and solve rate. We find the Tango* cost function catalyses strong improvements for the bidirectional DESP methods. Our method also achieves lower wall clock times while proposing synthetic routes of similar length, a common metric for route quality.
\end{abstract}

\section{Introduction}

Synthesis planning, where chemists design routes of chemical reactions to synthesise a complex molecule from simple or purchasable building blocks, is a key task in synthetic chemistry. The process used for this, \textit{retrosynthetic analysis}, involves recursively performing \textit{reversed} reactions, where a bond is broken to simplify a molecule into two or more component precursors \cite{corey1967general, Corey_Chelg_LogicofChemicalSynthesis}. Originally proposed by Corey in 1969, Computer-Assisted Synthesis Planning (CASP) aims to automate this process \cite{corey1969computer}. Since the seminal patent mining work of Lowe, which provided a large dataset of machine-readable chemical reactions, the CASP field has expanded significantly, with a plethora of approaches developed \cite{Lowe2017, Liu2017,Segler2017neural, coley2017computer, chen2020retro, tu2021permutation, Sacha2021molecule}. CASP systems typically have two primary components: a single-step retrosynthesis model, which decomposes a molecule into simpler precursors, and a search algorithm that explores the search graph constructed from outputs of the single-step model \cite{Segler2017neural, segler2018planning, schwaller2020predicting, chen2020retro, kishimotodepth, genheden2020aizynthfinder}. The iterative application of single-step models and exploration of the generated search space typically continues until a molecule is "solved," which is specified as having all leaf nodes belonging to a predefined set of purchasable building blocks. This approach of finding a path to any available precursor differs substantially from the approach expert chemists may take, where chemists can plan a synthesis with numerous constraints in mind, such as avoiding certain reactions and solvents, or starting from a specific precursor, known as a "structure-goal" \cite{Corey_Chelg_LogicofChemicalSynthesis}. By starting from a building block containing a key structural motif, the overall molecular complexity gain in a synthesis route can be lowered, a technique called "semi-synthesis" \cite{ojima1992new, brill2017navigating}. There is also considerable interest in repurposing waste compounds into useful products, a technique called "waste valorisation" \cite{wolos2022computer, lopez2024application, zkadlo2024computational}.

While designing constrained and steerable chemical synthesis is a daily practice in synthetic chemistry, it has received little attention in the CASP literature, with existing algorithms simply seeking to find any "valid" pathway to purchasable molecules. 

Recently, several approaches for \textit{starting material constrained} synthesis planning have been proposed with promising results \cite{johnson1992starting, yu2022grasp, yu2024double}. Existing solutions either rely on rule-based approaches or require complex systems with several interacting parts. In this work, we show that a \textit{general-purpose} and \textit{data-driven} retrosynthesis system can be adapted to \textit{starting material constrained} synthesis planning by the addition of a \textit{computed} node cost function. Our contribution is as follows:

\begin{enumerate}
        \item We use a computed node cost function, \textbf{TANimoto Group Overlap (}TANGO\textbf{)}, to guide the retrosynthetic search process towards enforced blocks. In this work, these blocks are limited to starting materials but could include key intermediates or molecular substructures.
        \item We show that by integrating TANGO into an existing \textit{general-purpose} search algorithm, we can tackle the constrained synthesis planning problem with comparable or superior results to existing, specialised methods.
        \item We further show that TANGO can be a drop-in replacement for retrosynthetic cost networks and demonstrate that by doing so, the performance of the recently proposed bidirectional search algorithm \citep{yu2024double} can be significantly improved.
        \item We compare the outputs of existing retrosynthetic value functions with the outputs of the TANGO node cost function and present a plausible explanation for the improved performance over existing starting material constrainted synthesis planning tools.
\end{enumerate}

\begin{figure}
    \centering
    \includegraphics[width=1.0\linewidth]{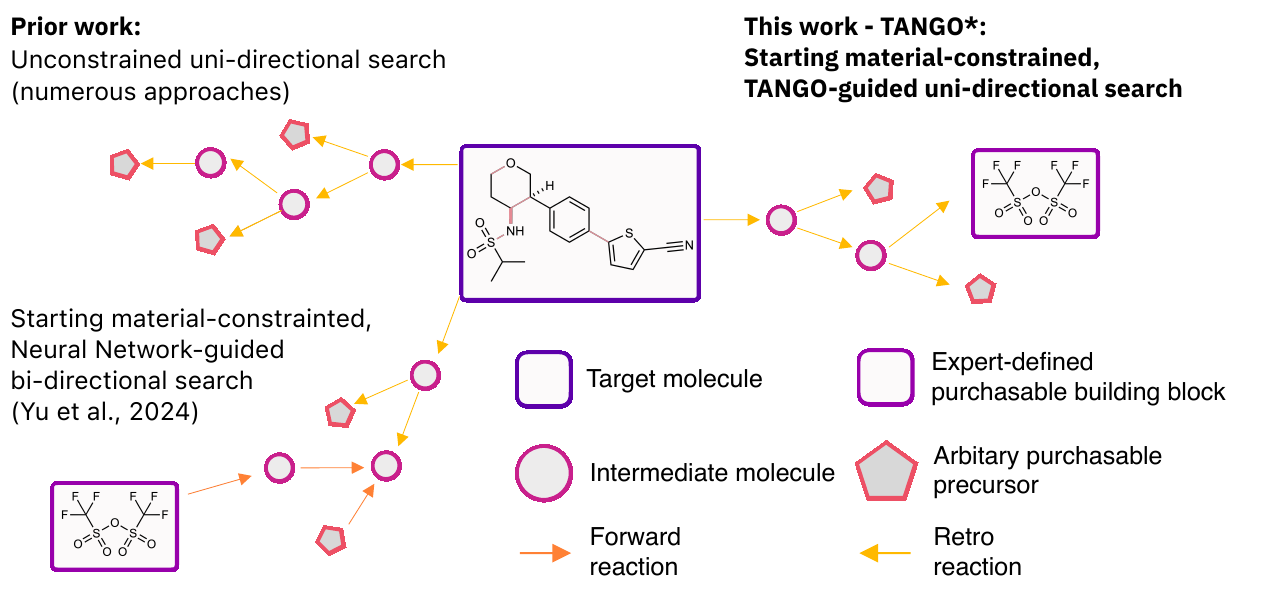}
    \caption{Comparison of existing constrained synthesis planning methods with Tango*}
    \label{fig:enter-label}
    \vspace{-7pt}
\end{figure}

\section{Related Work}

\textbf{Computer-Assisted Synthesis Planning (CASP)} tools typically formulate synthesis planning as a tree search, with each step corresponding to disconnecting a molecule into precursors through a retro'' chemical reaction. Two primary approaches are used for selecting retro'' reactions. Firstly, template-based methods extract chemical graph transformations from a corpus and train a neural network to select a transformation given an input~\cite{Segler2017neural,segler2018planning}. Template-free methods frame single-step retrosynthesis as a conditional language generation problem, with molecules encoded as SMILES strings or as a graph-edit prediction task~\cite{weininger1988smiles,schwaller2020predicting, Sacha2021molecule,zhong2023retrosynthesis}. Additionally, models that leverage graph features for direct generation have been developed~\cite{tu2021permutation,chen2023g}.
Significant focus has been placed on how to use single-step models in multi-step synthetic planning. Initial approaches used hand-curated rules, while more recent methods use neural-network guided graph exploration, such as Monte Carlo Tree Search (MCTS) or AND-OR graph search methods~\cite{pensak1977lhasa, szymkuc2016computer, segler2018planning, genheden2020aizynthfinder, kishimotodepth, chen2020retro}. A key development was driven by \citet{chen2020retro}, who proposed an A-star-like algorithm guided by a neural network that estimates the \textit{cost to synthesise} a molecule from any arbitrary purchasable building block~\cite{chen2020retro}. More novel methods have utilised self-play and experience-based learning to improve navigation of the search space~\cite{schreck2019learning,kim2021self, hong2023retrosynthetic,pdvn}. While single-step model performance continues to improve, this has not always been translated into the real-world performance of multi-step CASP systems \cite{thakkar2020datasets, schwaller2020predicting, maziarz2023re, modelsmatter}.

\textbf{Constrained Single-Step Retrosynthetic Models.} In recent years, there has been increased focus on introducing constraints into single-step retrosynthesis models with specific goals. Toniato \textit{et al.} utilised reaction class tokens to steer the output of single-step retrosynthetic transformers towards specific reaction classes \cite{toniato2023enhancing}. Following a similar approach, Thakkar \textit{et al.} introduced "disconnection prompts" to guide single-step models to break specific bonds \cite{thakkar2023unbiasing}. In the multi-step planning domain, Westerlund \textit{et al.} proposed a disconnection-aware transformer to encourage the breaking of bonds and allow the freezing of bonds during the search process, discarding any reaction that violates the frozen bond constraint \cite{westerlund2024constrained}. Interestingly, such bond constraints do not appear to impede the search process, indicating that simple, chemically informed rules can be powerful in data-driven retrosynthesis techniques.

\textbf{Starting Material Constrained Synthesis Planning.} Despite its potential use in waste valorisation and semi-synthesis, constrained synthesis planning has received limited attention in the literature. This approach imposes an additional constraint by focusing on the utilisation of specific starting materials. The LHASA program included such rules; however, they relied on expert-designed rules, limiting scalability \cite{johnson1992starting}. GRASP utilised reinforcement learning to develop a \textit{goal-driven} synthesis planning tool that can target either arbitrary products or specific starting materials \cite{yu2022grasp}. Recent state-of-the-art work proposed a bidirectional search algorithm, \textbf{Double-Ended Synthesis Planning (DESP)}, which uses both forward- and retro-expansion models, guided by a value network that estimates the cost of synthesising molecule \textit{m2} specifically from molecule \textit{m1} \cite{yu2024double}. Constrained synthesis planning has also emerged as a target in synthesisable molecular design. Guo et al. introduced a method for the \textit{de novo} generation of synthesisable molecules using enforced building blocks in the synthesis pathway \cite{guo2024takes}. To date, all starting material constrained synthesis planning tools have relied on specialised architectures, reinforcement learning, or expert-defined rules. In this work, we show instead that the problem can be approached with a simple cheminformatics calculation.

\begin{nolinenumbers} 
\noindent\makebox[\textwidth][c]{
    \begin{minipage}{0.80\textwidth}
    \begin{algorithm}[H]  % Reduce font size
        \caption{TANGO* Node Cost Function}
        \label{alg:tango-reward-calculation}
        \SetAlgoLined
        \DontPrintSemicolon
        \SetNoFillComment
        \SetSideCommentLeft
        \SetKwInOut{Input}{Input} % Properly formatted Input command
        \Input{ \\
           $MOL_{sm}$ \Comment{Enforced Starting Materials} \\
           $Node$ \Comment{Molecule Node} \\
           $k$ \Comment{Weight's influence of TANGO to node cost}\\
           $c$ \Comment{Scales TanSim vs. FMS} \\
           }
        \SetKwFunction{FMain}{Tango*NodeCost}
        \SetKwProg{Fn}{Function}{:}{}
        \Fn{\FMain{$Node$, $MOL_{sm}$}}{
            $reward_{node} \gets 0$\;
            \tcp{Loop through all specified starting materials}
            \ForEach{$mol_{sm} \in MOL_{sm}$}{
                \tcp{Compute reward for this starting material}
                $TanSim \gets \text{CalculateTanimotoSimilarity}(node, mol_{sm})$\;
                $FMS \gets \text{CalculateFMS}(node, mol_{sm})$\;
                $reward_{sm} \gets TanSim \cdot \textit{c}  +\, FMS \cdot (1 - c)$\;
                $reward_{node} \gets \max(reward_{node},\, reward_{sm})$\;
            }
            \KwRet{$k \cdot (1 - reward_{node})$ $+$ RetroCost}\;
        }
    \end{algorithm}% \end{wrapfigure}
    \vspace{-16pt}
    \end{minipage}
}
\vspace{-8pt}
\end{nolinenumbers} 
% \begin{nolinenumbers} 
% \begin{wrapfigure}{R}{0.55\textwidth}
% \begin{minipage}{0.50\textwidth}
% \begin{algorithm}[H]  % Reduce font size
%     \caption{TANGO* Node Cost Function}
%     \label{alg:tango-reward-calculation}
%     \SetAlgoLined
%     \DontPrintSemicolon
%     \SetNoFillComment
%     \SetSideCommentLeft
%     \SetKwInOut{Input}{Input} % Properly formatted Input command
%     \Input{ \\
%        $MOL_{sm}$ \Comment{Enforced Starting Materials} \\
%        $Node$ \Comment{Molecule Node}
%        }
%     \SetKwFunction{FMain}{Tango*NodeCost}
%     \SetKwProg{Fn}{Function}{:}{}
%     \Fn{\FMain{$Node$, $MOL_{sm}$}}{
%         $reward_{node} \gets 0$\;
%         \tcp{Loop through all specified starting materials}
%         \ForEach{$mol_{sm} \in MOL_{sm}$}{
%             \tcp{Compute reward for this starting material}
%             $TanSim \gets \text{CalculateTanimotoSimilarity}(node, mol_{sm})$\;
%             $FMS \gets \text{CalculateFMS}(node, mol_{sm})$\;
%             $reward_{sm} \gets TanSim \cdot \textit{c}  +\, FMS \cdot (1 - c)$\;
%             $reward_{node} \gets \max(reward_{node},\, reward_{sm})$\;
%         }
%         \KwRet{$k \cdot (1 - reward_{node})$ $+$ RetroCost}\;

% \end{algorithm}
% \end{minipage}
% \end{wrapfigure}   
% \end{nolinenumbers}
% \vspace{-4pt}
\section{Methods}

\textbf{TANGO* Overview}

We adopt the traditional structure of the synthesis planning problem using a top-down, or retrosynthetic, search algorithm. Due to its strong performance and built-in compatibility with node cost-based guidance functions, we use Retro* \citep{chen2020retro} as the baseline search algorithm. The Retro* algorithm uses a best-first approach, selecting the lowest-cost node at any given iteration. In the original paper, a neural network that \textit{estimates} the cost (number of chemical reactions) required to synthesise a given node from an arbitrary chemical building block is used to provide cost values for each node. To adapt the Retro* algorithm to the starting material constrained setting, we utilise a \textit{computed} function, \textbf{TANGO}, to estimate the structural and chemical similarity of a node to a \textit{specific} starting material. \textbf{TANGO} employs a weighted combination of Tanimoto Similarity and Fuzzy Matching Substructure (FMS) to compute molecular similarity. The Tango* cost is bounded by the interval ( [0, 1] ), while the Retro* cost is theoretically bounded by ( [0, $\infty$) ) and practically varies around the range ( [0, 10] ). To balance the influence of these mechanisms in the search dynamics, a scalar hyperparameter ( k ) is used to upweight the Tango* cost. The two cost functions are additively combined; further details are provided in Algorithm \ref{alg:tango-reward-calculation} and Appendix \ref{appendix:A2}.

\section{Experiments}

Tango* is built on the Retro* algorithm, using a reference implementation provided by DESP \cite{yu2024double}. For our estimation of node cost, we combine the original Retro* Value function \citep{chen2020retro} with an adaptation of the TANGO reward introduced by  \citet{guo2024takes}.

Our experiments are structured to answer the following questions: \textbf{(1)} Can a non-neural computed node cost function be used to adapt general-purpose synthesis planning tools to the constrained setting? \textbf{(2)} Can such a system outperform existing specialised models for starting material-constrained planning? \textbf{(3)} Can the cost function additionally provide improvements to existing bidirectional search methods? \textbf{(4)} As the TANGO function is empirically \textit{computed} as opposed to \textit{estimated}, does TANGO generalise from simple to harder datasets more effectively? 

\subsection{Experimental setup}
\textbf{Datasets.} To evaluate our system's performance, we utilise the common \textbf{USPTO-190} dataset introduced by \citet{chen2020retro}, which is a set of 190 challenging target molecules extracted from USPTO-Full. Additionally, we use the datasets introduced by \citet{yu2024double}, namely \textbf{Pistachio Reachable} and \textbf{Pistachio Hard}. The sets of \textit{target, starting material} pairs are extracted  for a set of commercial building blocks, we use canonical SMILES strings provided in the set of 23 million molecules from eMolecules used by \citet{chen2020retro} and \citet{yu2024double}. 

\textbf{Machine learning models.} To avoid variance due to subtle differences in data pre-processing techniques and to ensure a meaningful comparison, we use the Retro* value network and single-step retrosynthesis model provided by Yu et al.~\cite{yu2024double, maziarz2023re}. 

\textbf{Hyperparameter Optimisation.}
We use a hyperparameter \textit{k} to balance the starting material guidance of TANGO with the general guidance of the Retro* Value Network. To evaluate the ability of our method to generalise from simpler to more complex molecules, we choose the Pistachio Reachable dataset for hyperparameter tuning. We find a value of \textit{k} = 25 optimises both Solve Rate and Average Number of expansions. We employ an additional parameter, \textit{c}, to specify the ratio of FMS to Tanimoto Similarity, with \textit{c} defining the FMS weight. Through empirical testing, we determine the optimal value to be \textit{c} = 0.3. In the results section, we will refer to Tango with \textit{c} = 0.0 as Tango(1, 0) and Tango with \textit{c} = 0.3 as Tango(0.7, 0.3).

\begin{table}
% \vspace{-pt}
  \caption{Summary comparison between baseline methods and Tango* across the three benchmarks. Baseline results for Retro*, GRASP,Retro* + D and the DESP methods are taken from \citet{yu2024double}. Solve rate is the fraction of \textit{(target, starting material)} pairs solved within the expansion limit. Tango DESP methods use a (1, 0) weighting of Tanimoto Similarity to FMS. Best overall results are in \textbf{bold} and best uni-directional results are \underline{underlined}}.
  \label{primary-results}
  \centering
  \small
  \begin{tabular}{lcccccccccccc}
    \toprule
    Algorithm & \multicolumn{4}{c}{USPTO-190} & \multicolumn{4}{c}{Pistachio Reachable} & \multicolumn{4}{c}{Pistachio Hard} \\
    \cmidrule(lr){2-5} \cmidrule(lr){6-9} \cmidrule(lr){10-13}
    & \multicolumn{3}{c}{Solve Rate (\%) $\uparrow$} & $\overline{N}$ $\downarrow$ &  \multicolumn{3}{c}{Solve Rate (\%) $\uparrow$} & $\overline{N}$ $\downarrow$ &  \multicolumn{3}{c}{Solve Rate (\%) $\uparrow$} & $\overline{N}$ $\downarrow$ \\ Expansion Budget
    & 100 & 300 & 500 & & 50 & 100 & 300 & & 100 & 300 & 500 &  \\
    \midrule
    Random & 4.2 & 4.7 & 4.7 & 479 & 16.0 & 26.7 & 40.7 & 325 & 6.0 & 12.0 & 13.0 & 452 \\
    BFS & 12.1 & 20.0 & 24.2 & 413 & 48.7 & 57.3 & 74.0 & 169 & 16.0 & 26.0 & 29.0 & 390 \\
    MCTS & 20.5 & 32.1 & 35.3 & 364 & 52.0 & 72.7 & 85.3 & 111 & 27.0 & 31.0 & 32.0 & 361 \\
    Retro* & 25.8 & 33.2 & 35.8 & 351 & 70.7 & 78.0 & 92.7 & 73 & 32.0 & 35.0 & 37.0 & 342 \\
    GRASP & 15.3 & 21.1 & 23.7 & 410 & 46.7 & 51.3 & 66.7 & 198 & 14.0 & 22.0 & 29.0 & 402 \\
    Retro*+D & 27.4 & 32.6 & 37.4 & 348 & 77.3 & 87.3 & 96.0 & 49 & 31.0 & 40.0 & 42.0 & 323 \\
    DESP-F2E & 30.0 & 35.3 & 39.5 & 340 & 84.0 & 90.0 & 96.0 & 41 & 35.0 & 44.0 & 50.0 & 300 \\
    DESP-F2F & 29.5 & 34.2 & 39.5 & 336 & 84.5 & 88.9 & 97.3 & 38 & 39.0 & 45.0 & 48.0 & 293 \\
    \midrule
    \textbf{Ours} \\
    Tango(1, 0)* & \textbf{\underline{36.3}} & \underline{41.1} & \underline{42.6} & \underline{313} & \underline{84.5} & \underline{90.6} & \underline{97.3} & \underline{32} & \underline{40.0} & \underline{45.0} & \underline{47.0} & \underline{290} \\
    % Tango(0.7, 0.3)* & 35.7 & 40.5 & 42.1 & 316 & 86.7 & 94.0 & 98.0 & 29 & 39.0 & 44.0 & 46.0 & 295 \\
    Tango-F2E & 33.1 & 40.0 & 41.5 & 317 & 88.7 & 92.0 & 98.7 & 27 & 39.0 & 45.0 & 49.0 & 290 \\
    Tango-F2F & 33.2 & \textbf{45.3} & \textbf{53.7} & \textbf{291} & \textbf{91.3} & \textbf{95.3} & \textbf{99.3} & \textbf{18} & \textbf{47.0} & \textbf{59.0} & \textbf{63.0} & \textbf{231} \\
    \bottomrule
  \end{tabular}
  \vspace{-8pt}
\end{table}

\subsection{Tango* solve rate and algorithm efficiency}

While there remains a lack of a clearly agreed-upon 'gold standard' for reporting the quality of synthetic routes generated by CASP systems, a few metrics are commonly used. The \textbf{Solve Rate} indicates the system's ability to find viable solutions, while the \textbf{Average Route Length} (in terms of the number of reactions) indicates route quality. In addition to reporting metrics regarding the outputs of Tango*, we aim to evaluate how efficient Tango* is at navigating the retrosynthetic search space and assess the effect of computational overhead on the system. To this end, we report the \textbf{Average Number of Expansions} (\(\overline{N}\)) and \textbf{Wall Clock Time}. 

The primary results are displayed in Table \ref{primary-results}. Tango(1, 0)* demonstrates improvements in the starting material-constrained setting and consistently outperforms the neural network-enhanced Retro* (referred to as Retro* + D) across all benchmarks and expansion limits while displaying greater algorithmic efficiency, as measured by the average number of expansions \(\overline{N}\). In addition, Tango(1, 0)* achieves higher or comparable solve rates to both DESP methods across all three datasets, doing so with a strictly lower average number of expansion calls, clearly demonstrating the power of the TANGO reward to navigate the retrosynthetic action space. We find that the value of \textit{k}, optimised on Pistachio Reachable, shows strong generalisation performance to the more challenging datasets, with Tango* being the best-performing method across all iteration levels on USPTO-190. 

\begin{table}
\caption{Inference time and Mean Solved Route Length for the evaluated methods. Route length comparisons are made on the routes solved by all methods.}
\small
\label{inference-times}
  \begin{tabular}{lccccccc}
    \toprule
    Algorithm & \multicolumn{2}{c}{USPTO-190} & \multicolumn{2}{c}{Pistachio Reachable}
    & \multicolumn{2}{c}{Pistachio Hard} \\
    \cmidrule(lr){2-3} \cmidrule(lr){4-5} \cmidrule(lr){6-7} 
    & \shortstack{Route Length \\ (61 Routes)} & \shortstack{Wall Clock \\ Time (s)} & \shortstack{Route Length \\ (114 Routes)} &  \shortstack{Wall Clock \\ Time (s)} & \shortstack{Route Length \\ (36 Routes)} &  \shortstack{Wall Clock \\ Time (s)} \\
    Retro* & 5.30 & 58.1 & 4.64 & 10.2 & 4.67 & 56.3 \\
    Retro* + D & 5.56 & 64.1 & 4.67 & 8.3 & 4.67 & 55.2 \\
    DESP-F2E & 5.13 & 66.5 & 4.51 & 8.6 & 4.56 & 56.3 \\
    DESP-F2F & 5.51 & 109.4 & 4.46 & 8.2 & 4.44 & 61.8 \\
    \midrule
    Tango(1, 0)* & 5.06 & \textbf{55.8} & 4.56 & 5.8 & 4.67 & \textbf{47.5} \\
    Tango-F2E & \textbf{4.44} & 75.3 & 4.24 & 6.5 & \textbf{4.29} & 54.2\\
    Tango-F2F & 5.06 & 146.4 & \textbf{4.04} & \textbf{4.9} & 4.40 & 72.8 \\    
    \bottomrule
\end{tabular}
\vspace{-8pt}
\end{table}

We perform an ablation of the FMS : Tanimoto Similarity weighting in the \textbf{TANGO} cost function, which we refer to as Tango(0.7, 0.3). We find that although the incorporation of FMS into the cost function improves the solve rate and reduces expansion calls for Pistachio Reachable, such results do not carry over to the more challenging datasets. We hypothesise that Tanimoto Similarity offers greater granularity for guidance than FMS, enabling higher performance on more complex datasets. 

To examine the general applicability of the TANGO cost in guiding various search algorithms, we explore its integration into the recently proposed bidirectional search methods, DESP-F2F and DESP-F2E \citep{yu2024double}. This integration is achieved by replacing the pairwise synthetic distance network, \textit{D}, with the TANGO cost function. The hyperparameters \textit{k} and \textit{c} are set to the same values as Tango (1,0). Our findings show that the addition of TANGO reward generally leads to a substantial increase in the solve rate for both DESP methods, while also reducing the average number of expansions and route length. Particularly noteworthy is the impressive performance of Tango-F2F at high expansion budgets, where it achieves a 99.3\% solve rate on Pistachio Reachable and improves accuracy by approximately 25\% compared to the next best method on the more challenging UPSTO-190 and Pistachio Hard datasets. We note that as the added bidirectional search of DESP outperforms Retro* + \textit{D}, TANGO-DESP methods should be expected to outperform Tango*.
\begin{figure}

    \centering
    \includegraphics[width=0.95\linewidth]{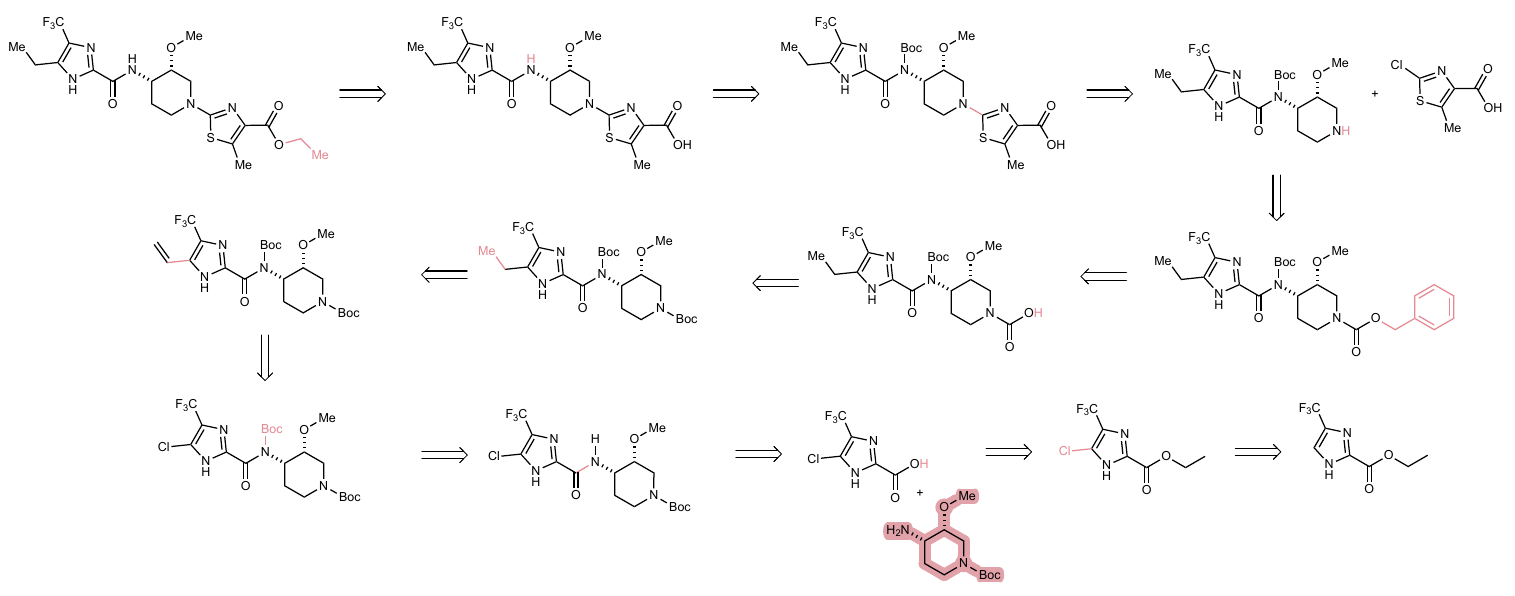}
    \caption{Here we demonstrate a meaningful 12-step route generated by our method on a \textit{(target, starting material)} pair not solved by the best performing DESP \citep{yu2024double} method. Constrained starting material highlighted in red; bonds/atoms disconnected shown in red.}
    \label{fig:route145}
    \vspace{-8pt}
\end{figure}

\subsection{Wall clock time and route length ablations}

As both DESP and Tango* introduce computational overhead that may decouple the number of expansion calls from the computational resources required, we report the wall clock time. Tango* consistently achieves a lower wall clock time than alternative starting material-constrained methods. Finally, we investigate the average number of reactions per solved route for each method. Tango* achieves shorter route lengths than all existing methods on USPTO-190, but only matches existing methods on other datasets. The strongest results in terms of route length come from the combination of TANGO with bidirectional search methods F2E and F2F, with one of them displaying the shortest routes for all of the datasets. This result is revealing; existing methods use a neural network directly trained to predict synthetic distance, and yet it fails to provide significantly stronger guidance towards shorter synthetic routes than a simply molecular similarity measure. This leads to the obvious question, just how effective are such neural networks at estimating the synthetic distance of a node, and how reliable is this estimation at test time?

\subsection{Why does Tango* work?}

Despite access to a \textit{starting material constrained} node cost function (one with access to information from ground truth routes at test time), Retro* +\textit{D} does not show a substantial increase in performance compared to Retro*. In contrast, Tango*'s incorporation of a privileged node cost function provides significant performance improvements. We hypothesize that, as a computed cost function, the TANGO cost function should be relatively invariant to the molecular inputs and maintain strong performance at test time.
Let $V: M \rightarrow \mathbb{R}$ be a node cost function where M is the space of molecules. For node cost to effectively guide retrosynthetic search the function should $generally$ satisfy:
\begin{equation}
V(m_i) > V(m_{i+1}) \quad \forall i \in {1,...,n-1}
\end{equation}
where $m_1,...,m_n$ represents molecules along a synthetic path from root to starting material. 

A potential caveat to the above constraint might be "tactical combinations", the synthetic complexity of a molecule temporarily increases during a retrosynthesis, to allow a major complexity-reducing reaction \citet{gajewska_algorithmic_2020}. We believe that retrosynthetic value functions should continue to show synthetic distance estimate decreases even for the complexity-increasing step.
We also expect a stronger monotonic decrease for routes that are \textit{solved} by a method, compared to routes \textit{not solved}, as this indicates that the node cost function can accurately prioritise nodes in the search tree.  To systematically evaluate these hypotheses about TANGO's effectiveness and empirically assess the relative strength of different guidance functions, we analyse their behavior on ground truth synthetic routes in the test set. Using the USPTO-190 dataset, we extract the linear synthetic path from root molecule $m_r$ to expert-defined starting material $m_s$. We define synthetic distance $d(m_1, m_2)$ as the minimum number of reactions required to synthesize $m_1$ from $m_2$. For each molecule $m_i$ in this path, we calculate:
\begin{itemize}
    \item $d(m_i, m_s)$: Ground truth synthetic distance
    \item $D(m_i, m_s)$: Neural network estimation of synthetic distance
    \item $T(m_i, m_s)$:      TANGO cost molecular similarity
\end{itemize}

To isolate the cost function, we fix the search algorithm, in this case focusing on the Retro* algorithm with either TANGO or \textit{D} as a starting material-guided cost function. We then take the 4 sets of routes that are solved and not solved by Tango* and Retro* +\textit{D}. We plot the corresponding starting material constrained cost function, TANGO and \textit{D} respectively, as costs for each node in the ground truth synthetic routes.

\begin{figure}
    \centering

    \includegraphics[width=1.0\linewidth]{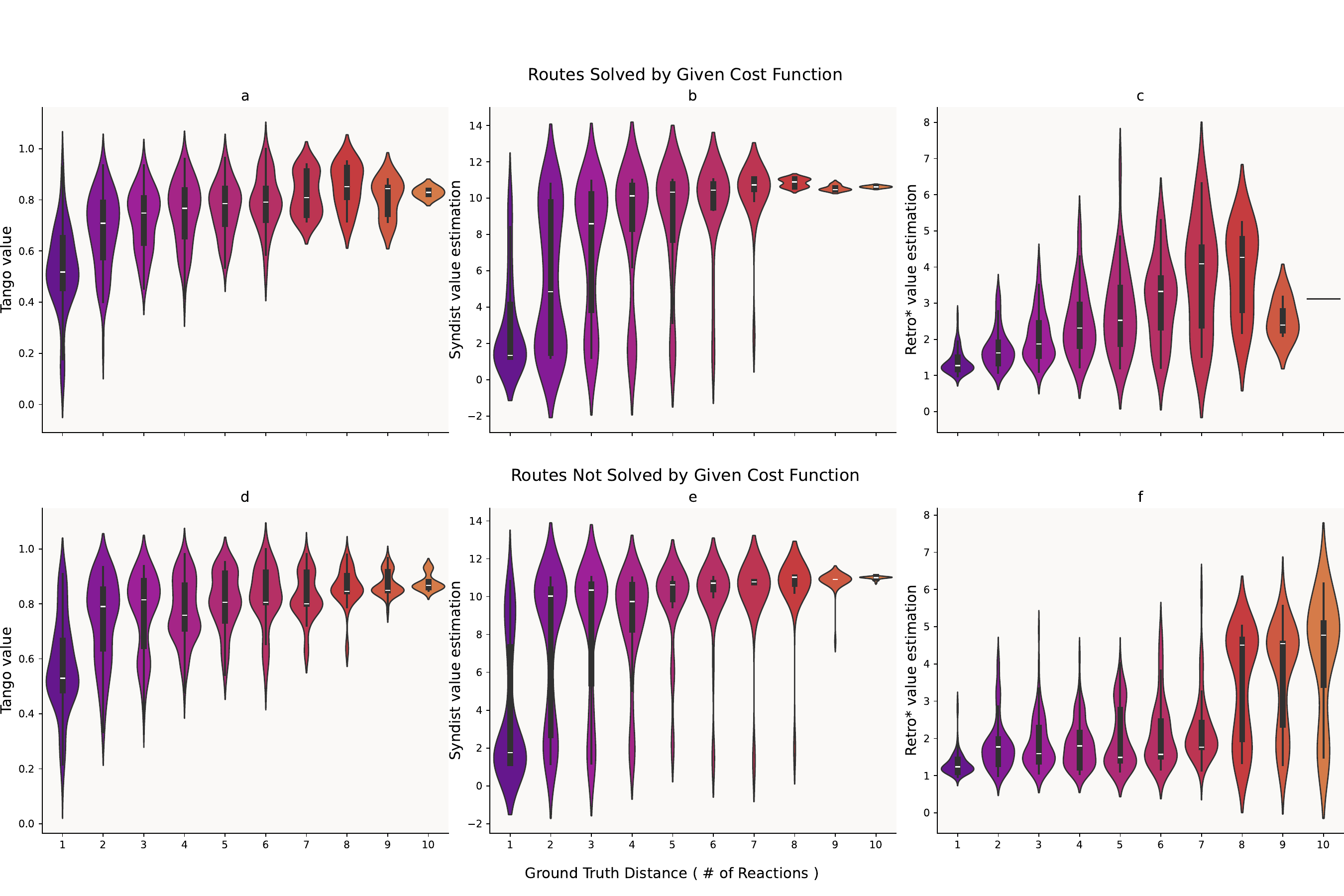}
    \caption{A comparison of node cost estimates for USPTO-190 routes solved and not solved by Retro* search using the corresponding cost function.\textbf{(a)} Tango Cost for routes solved by Tango*, \textbf{(b)} SynDist Cost for routes solved by Retro* + \textit{D}, \textbf{(c)} Retro* cost function  estimates for routes solved by Retro*, \textbf{(d)} Tango Cost for routes not solved by Tango*, \textbf{(e)} SynDist cost for routes not solved by Retro* + \textit{D} and \textbf{(f)} Retro* cost for routes not solved by Retro*}
    \label{fig:reward_func_plots}
    \vspace{-16pt}
\end{figure}

We show the results of this experiment in Figure \ref{fig:reward_func_plots}. The neural network estimated synthetic distance \textbf{(b)} and \textbf{(e)} displays an unexpected bimodal distribution, with peaks at low and high estimates. It displays consistently high absolute error and is unable to provide a granular, monotonically decreasing estimate of synthetic distance. For routes that Retro* + \textit{D} solves, the mean synthetic distance estimate begins to decrease when the ground truth distance is under two but still displays substantial error margins. This is exacerbated for routes that Retro* does not solve, with the synthetic distance estimation consistently varying around 10 if the ground truth distance is greater than 1. We note this fits with the training strategy of \textit{D} described in \cite{yu2024double}, which augments the training set with synthetic "negative samples" of $(target, starting material)$ pairs. These samples are generated by selecting two molecules that are disconnected in the directed graphs formed by linking reactants and products in USPTO, and are assigned a fixed "distance" value of $10$.
While such a method should be robust in a \textit{dense} dataset where all theoretically synthesisable molecules are interconnected, USPTO is notably \textit{sparse}. Consequently, many of these "disconnected" samples may, in fact, be feasibly synthesisable. Furthermore, the constraint enforcing a Tanimoto Similarity of less than 0.7 may be overly permissive. We observe that the majority of molecules within the same synthetic route typically exhibit Tanimoto Similarities far lower than 0.7, suggesting that this sampling approach may introduce significant noise into the training process.
% We hypothesise that this training and sample selection method for \textit{D} may effectively turn synthetic distance \textit{estimation} into a binary classification of "close" or "distant" based on Tanimoto similarity between the pair of molecules. 

In comparison, the computed TANGO cost function \textbf{(a)} and \textbf{(d)} exhibits a clear and granular monotonic decrease across routes solved by Tango*. On routes not solved by Tango*, the monotonic decrease is less pronounced but still present. We hypothesize that this strong granularity and monotonicity enables TANGO cost-guided search algorithms to achieve substantially improved solve rates compared to neurally guided methods. Based on this observation, we investigated whether similar findings apply to \textit{non-privileged} (non-goal oriented) neural guidance functions, such as the original Retro* cost function. Prior work has shown that omission of the Retro* cost function has limited effects on retrosynthetic search performance, with both \citet{chen2020retro} and \citet{maziarz2023re} finding that setting the value to a constant resulted in only minor performance changes on USPTO-190. The results for solved and unsolved routes are shown in Figure \ref{fig:reward_func_plots} \textbf{(c)} and \textbf{(f)}, respectively. In contrast to the plots for the \textit{D} cost function, which consistently \textit{overestimates} cost, the Retro* cost function consistently \textit{underestimates} it. While it displays clear monotonicity on solved routes, this pattern does not extend to unsolved routes, where the cost function hovers around 1-2 before splitting into a bimodal distribution with \textit{lower error} at ground truth distances greater than 7. This is likely a function of the power-law like distribution of reaction sequence lengths in USPTO, with the majority of synthetic routes consisting of only a single step. These results indicate that value networks trained exclusively on positive samples using simple training methods such as MSE and Consistency loss perform substantially better at estimating synthetic distance. \textit{unprivileged} setting.
\begin{figure}
    \centering
    \includegraphics[width=0.90\linewidth]{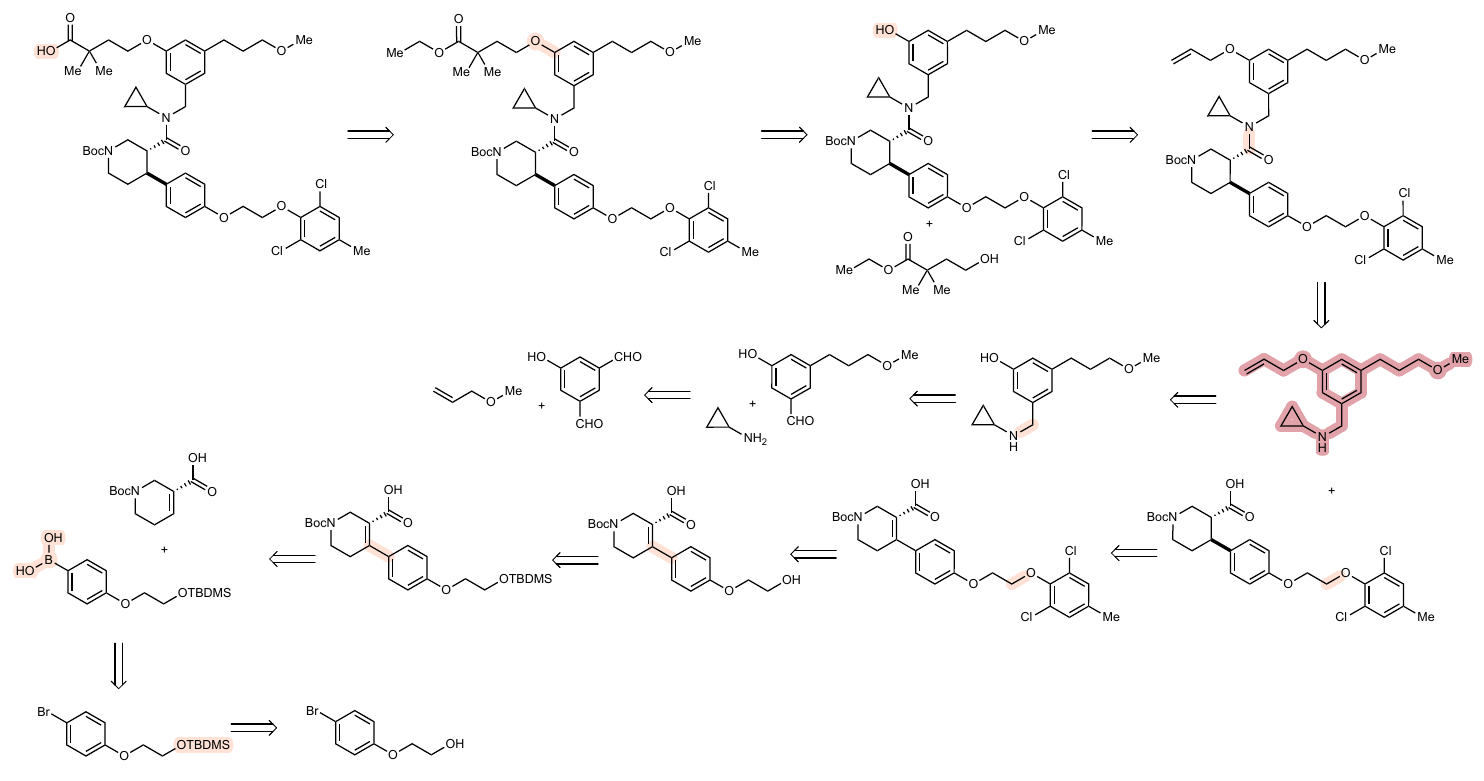}
    \caption{Here we demonstrate a feasible 10-step route generated by Tango-DESP-F2F on a \textit{(target, starting material)} pair not solved by the neural guided DESP-F2F method. Constrained starting material is highlighted in red; bonds/atoms disconnected are shown in red.}
    \label{fig:solved_tango_not_desp}
\end{figure}
\vspace{-8pt}
\begin{figure}[h]
    \centering
    \includegraphics[width=0.80\linewidth]{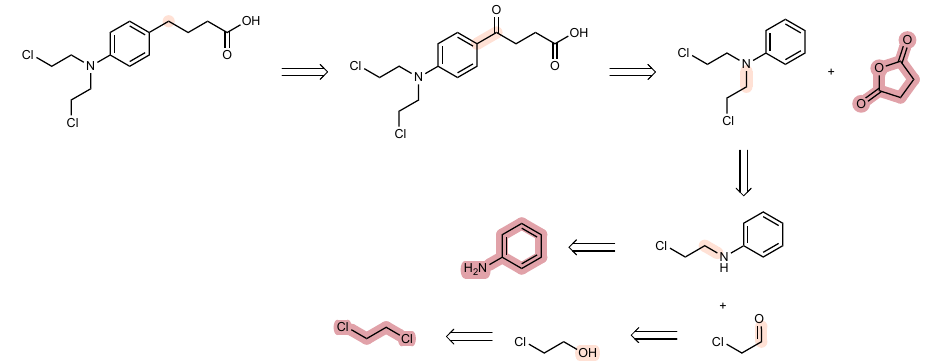}
    \caption{Here we show a feasible synthesis route to the chemotherapy drug, Chlorambucil, a WHO essential medicine, synthesised entirely from renewable or industrial waste feedstocks.}
    \label{fig:chlorambucil}
\end{figure}
\vspace{-6pt}
\subsection{Case Study: Synthesis of Useful Compounds from Renewable/Waste Feedstock}

A key aim of starting-material-constrained synthesis planning is to enable the discovery of synthetic pathways to useful compounds from renewable or waste feedstocks. Previous evaluations relied on a database of 17 million chemical building blocks from the eMolecules database. We aim to demonstrate the effectiveness of our method, Tango*, in finding synthesis pathways to useful small molecules starting \textit{exclusively} from renewable or waste feedstocks. For renewable building blocks, we use a set of 146 small molecules previously curated by \citet{wolos2022computer}. For useful compounds, we extract a set of 110 small molecules from a curation of the WHO List of Essential Medicines, previously developed by \citet{gao2020combining}. We conduct the search using the Tango(1, 0) set of hyperparameters previously described, but set the expansion budget to 1,000 model calls. In Figure \ref{fig:chlorambucil}, we present a strong route, discovered by Tango* but not Retro*, to the chemotherapy drug Chlorambucil, starting exclusively from renewable starting materials. We find all proposed reactions and the complete synthesis are directly reported in the literature \cite{yang2020iridium, Gangjee2009Synthesis}

\section{Conclusion}

In this work, we introduce Tango*, a simple adaptation of the Retro* algorithm to the starting material-constrained setting without any model retraining. We demonstrate that our \textbf{TANGO} guided search method strictly outperforms the similar neural network-guided Retro* + \textit{D}. Despite relying on single-ended search, Tango* either outperforms or matches the performance of specialised DESP models and search algorithms, providing routes that satisfy the specified goal for a greater number of compounds. Application of the TANGO node cost function to the DESP methods also yields substantial improvements, particularly to the F2F method, which achieves the strongest solve rate performance of all investigated systems. It proposes routes with a comparable length and does so with a lower number of expansion calls and reduced wall clock time. 

We show that existing neural node cost functions fail to provide a granular and monotonic decrease in node cost throughout a retrosynthesis pathway, particularly struggling on more challenging routes. In contrast, the computed Tango* cost function displays better monotonicity and granularity on both solved and unsolved routes. This work indicates that there may be substantial room for improvement in developing novel guidance functions for retrosynthesis tools.

We anticipate that future developments in similar methods will unlock synthesis planning tools with diverse and flexible structure constraints, allowing expert chemists to specify key intermediates or predefined substructure goals at any position in the synthetic route.

\textbf{Code and Data availability}
The code used to produce these results is available here \url{https://github.com/schwallergroup/TangoStar}
\begin{ack}
This research was funded by the Swiss National Science Foundation (SNSF) [214915]. Additional funding was provided by NCCR Catalysis (grant number 225147), a National Centre of Competence in Research funded by the Swiss National Science Foundation.

\end{ack}
\newpage
\bibliographystyle{unsrtnat}
\bibliography{bib}

\medskip

%%%%%%%%%%%%%%%%%%%%%%%%%%%%%%%%%%%%%%%%%%%%%%%%%%%%%%%%%%%%
\newpage
\appendix

\section{Appendix}

\textbf{A.1: Compute Details}

Wall clock time comparisons were implemented on a GPU-enabled workstation with the following specifications.
\begin{itemize}

    \item CPU: 12-core AMD Rysen 9 7900X
    \item RAM Memory : 64 GB
    \item GPU : NVIDIA A6000 48 GB
\end{itemize}

\textbf{A.2: Retro* Algorithm}\label{appendix:A2} 

We note this algorithm description is largely taken from the DESP paper \citet{yu2024double}.

Retro* defines the following quantities:
\begin{itemize}

\item \( V_m \): For a molecule \( m \), \( V_m \) is an unconditional estimate of the minimum cost required to synthesise \( m \). This estimate is provided by a neural network.

\item \( \text{rn}(m|G) \): For a molecule \( m \) in the search graph \( G \), the "reaction number" \( \text{rn}(m|G) \) represents the estimated minimal cost to synthesise \( m \).

\item \( V_t(m|G) \): For a molecule \( m \) in the search graph \( G \) with the target molecule \( p^* \), \( V_t(m|G) \) denotes the estimated minimal cost to synthesise \( p^* \) starting from \( m \).
\end{itemize}
Retro* operates through iterative phases of selection, expansion, and update. We follow the DESP implementation of Retro* as follows:

**Selection:** Choose the molecule from the set of frontier nodes \( F \) that minimises the expected cost of synthesising the target \( p^* \) given the current search graph \( G \):

\begin{equation}
m_{\text{select}} = \arg\min_{m \in F} V_t(m|G)
\end{equation}

**Expansion:** As detailed in Algorithm 2, apply a one-step retrosynthesis model to the selected node \( m_{\text{select}} \), and add the resulting reactions and precursor molecules to the search graph \( G \). Initialise each new molecule node with:

\[
\text{rn}(m|G) \leftarrow V_m
\]

**Update:** 

First, propagate the reaction number values upward to ancestor nodes. For a reaction node \( R \), update its reaction number as the sum of its childrens' reaction numbers plus the cost of the reaction \( c(R) \):

\begin{equation}
\text{rn}(R|G) \leftarrow c(R) + \sum_{m \in \text{ch}(R)} \text{rn}(m|G)
\end{equation}

For a molecule node \( m \), update its reaction number to be the minimum reaction number among its child reactions:

\begin{equation}
\text{rn}(m|G) \leftarrow \min_{R \in \text{ch}(m)} \text{rn}(R|G)
\end{equation}

Next, propagate the values of \( V_t(m|G) \) downward to descendant nodes. Starting from the target molecule \( p^* \):

\begin{equation}
V_t(p^*|G) \leftarrow \text{rn}(p^*|G)
\end{equation}

For subsequent reaction nodes \( R \), update the value as:

\begin{equation}
V_t(R|G) \leftarrow \text{rn}(R|G) - \text{rn}(\text{pr}(R)|G) + V_t(\text{pr}(R)|G)
\end{equation}

Finally, for molecule nodes \( m \) that are not the target \( p^* \):

\begin{equation}
V_t(m|G) \leftarrow \min_{R \in \text{pr}(m)} V_t(R|G)
\end{equation}

Here, \( \text{ch}(R) \) denotes the set of child molecules of reaction \( R \), and \( \text{ch}(m) \) represents the set of child reactions of molecule \( m \). Similarly, \( \text{pr}(R) \) denotes the parent molecule of reaction \( R \), and \( \text{pr}(m) \) represents the set of parent reactions of molecule \( m \).

This implementation ensures that at each iteration, the algorithm selects the most promising node to expand based on the estimated cost, propagates cost updates throughout the search graph, and efficiently guides the search towards the most cost-effective synthesis pathways.

\textbf{A.3: Dataset Details}\label{appendix:A3}

We provide details on the used datasets in Table \ref{table:table_datasets}. The Table is taken directly from the original paper \cite{yu2024double}.

\begin{table}[h]

\centering
\caption{Benchmark dataset summary. Avg. In-Dist. \% is the mean percentage of reactions in each route within the top 50 suggestions of the retro model. Unique Rxn.\% is the ratio of deduplicated reactions to total reactions across all routes. Avg. \# Rxns. is the mean number of reactions in each route, and Avg. Depth is the mean number of reactions in the longest path of each route.}
\label{table:table_datasets}
\BlankLine
\begin{tabular}{lccccc}
\hline
Dataset & \# Routes & Avg. In-Dist. \% & Unique Rxn. \% & Avg. \# Rxns. & Avg. Depth \\
\hline
USPTO-190 & 190 & 65.6 & 50.5 & 6.7 & 6.0 \\
Pistachio Reachable & 150 & 100 & 86.1 & 5.5 & 5.4 \\
Pistachio Hard & 100 & 59.9 & 95.2 & 7.5 & 7.2 \\
\hline
\end{tabular}
\end{table}

\textbf{A.4: TANGO Hyperparameter Screening}\label{appendix:A4}

Here we provide a hyperparameter screen of TANGO weight and Tanimoto weight across the 3 Tango augmented methods, \textbf{Tango*}, \textbf{TANGO DESP-F2E} and \textbf{TANGO DESP-F2F}. Due to computational resource limitations, this screen was conducted exclusively on Pistachio Reachable with a small expansion budget of 50. In general, a higher Tango Weight (> 15) tends to lead to a higher solve rate, consistent accross all methods.

A Tanimoto weight of 0.75 performs strongly for the \textbf{TANGO DESP-F2F} \textbf{Tango*} methods on the Pistachio Reachable test set, consistent with previous results.

\begin{figure}[h]
    \centering
    \includegraphics[width=0.95\linewidth]{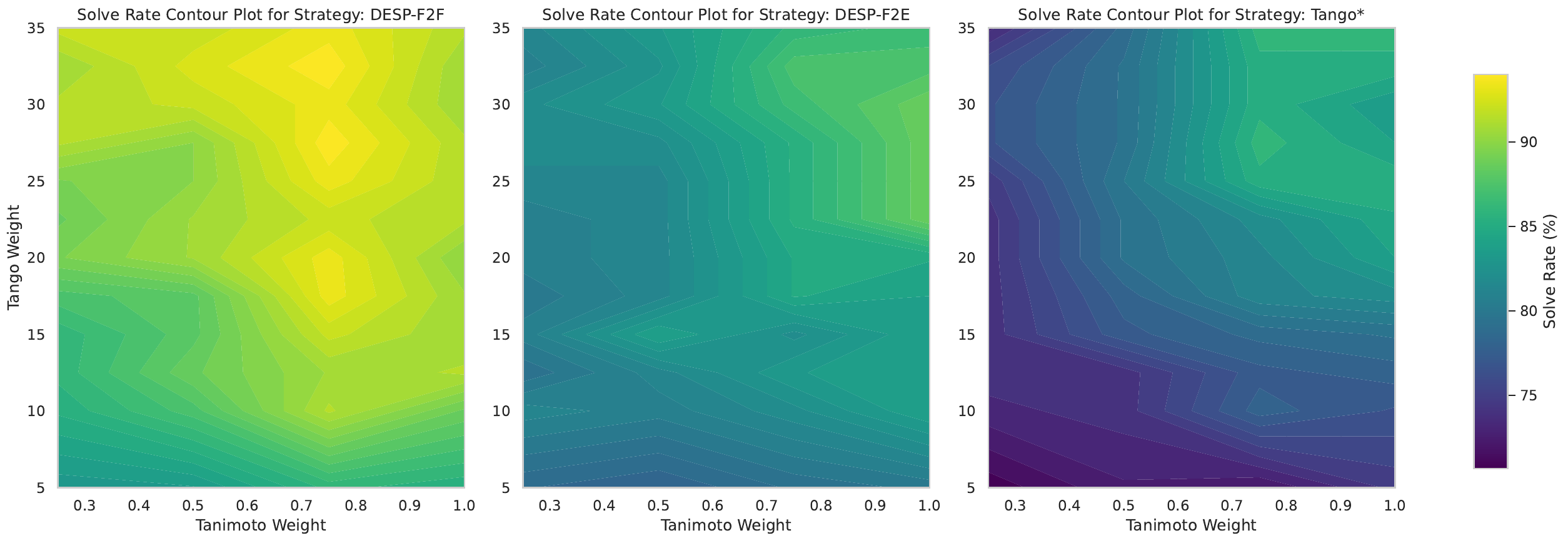}
    \caption{A hyper-parameter screen conducted on Pistachio Reachable with an expansion limit of 50.}
    \label{fig:gridsearch}
\end{figure}

\textbf{A.5: Renewable synthesis of WHO Essential Medicines datasets}

The sources of small molecule WHO Essential Medicines and renewable and waste compound feeds\citet{gao2020combining, wolos2022computer} are not machine-readable. We used a vision-enabled large language model (GPT-4o) to parse compound names from the Supplementary Information sections of the aforementioned papers.

\textbf{A.6: Value function failiure points}

In Figure 7, we analyze discrepancies between model-predicted number of synthesis steps (Retro* and SynthDist) and ground truth values across four synthetic transformations. In these examples, models overestimate synthetic complexity compared to ground truth, likely due to their focus on structural features rather than strategic synthetic planning. This is evident in: (a) ring-opening of tetrahydropyran introducing rotatable bonds of an alkyl chain, (b) TBDMS protection adding heavy atoms, (c) oxidation of primary alcohols changing atom connectivity and adding heavy atoms, and (d) lactone formation creating additional ring complexity. These cases demonstrate how the models may misinterpret strategic intermediates with increased synthetic complexity as being further away from commercially available building blocks while in reality, these intermediates help disconnections over multiple steps, suggesting potential areas for improvement in retrosynthetic prediction algorithms.

\begin{figure}[h]
    \centering
    \includegraphics[width=0.95\linewidth]{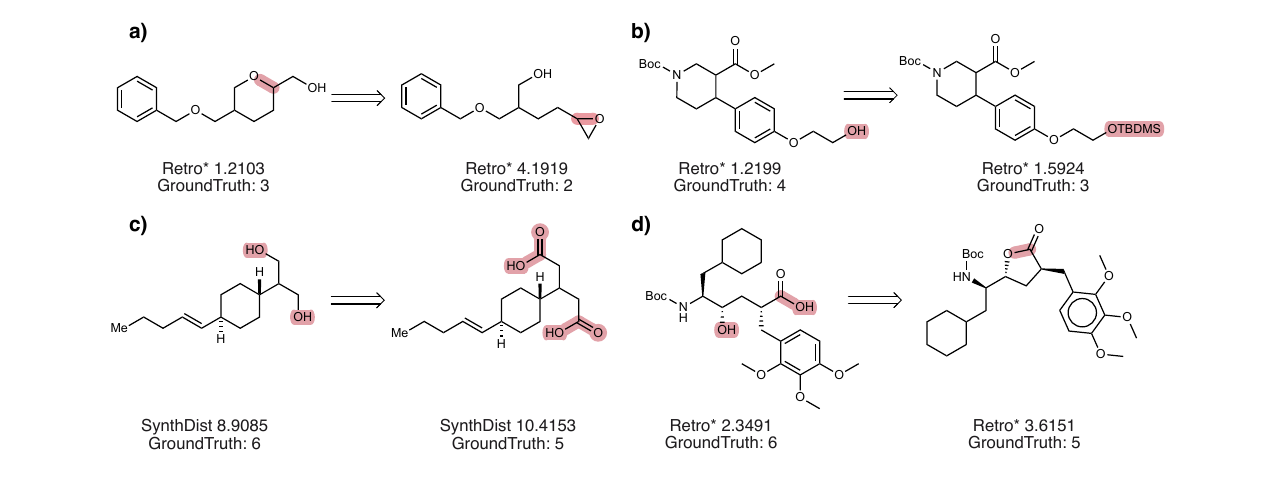}
    \caption{Failure points of value networks (Retro* and SynDist) compared with ground truth synthetic distance values. Bonds and atoms that are being modified are highlighted.} 
    \label{fig:gridsearch}
\end{figure}

\end{document}